\documentclass[%
preprint,
superscriptaddress,
preprintnumbers,
footinbib,
amsmath,amssymb,
aps,prfluids,
floatfix,
longbibliography
]{revtex4-2}

\usepackage[subrefformat=parens,labelformat=parens,caption=false]{subfig} 
\usepackage{bm,color,graphicx}

\allowdisplaybreaks

\usepackage{hyperref}
\hypersetup{
  colorlinks = true,
  linkcolor  = blue,
  citecolor  = blue,
  urlcolor   = blue,
  breaklinks = true
}

\begin{document}

\title{Shape of spreading and leveling gravity currents in a Hele-Shaw cell with flow-wise width variation}

\author{Zhong Zheng}
\email[]{zzheng@alumni.princeton.edu, zhongzheng@sjtu.edu.cn}
\affiliation{State Key Laboratory of Ocean Engineering, Shanghai, China, 200240}
\affiliation{School of Naval Architecture, Ocean and Civil Engineering, Shanghai Jiao Tong University, Shanghai, China, 200240}

\author{Aditya A.\ Ghodgaonkar}
\email[]{adighod@mit.edu}
\affiliation{School of Mechanical Engineering, Purdue University, West Lafayette, IN 47907, USA}
\affiliation{Department of Mechanical Engineering, Massachusetts Institute of Technology, Cambridge, MA 02139, USA}

\author{Ivan C.\ Christov}
\email[]{christov@purdue.edu}
\homepage[]{http://christov.tmnt-lab.org}
\thanks{Corresponding author.}
\affiliation{School of Mechanical Engineering, Purdue University, West Lafayette, IN 47907, USA}


\begin{abstract}
We study the spreading and leveling of a gravity current in a Hele-Shaw cell with flow-wise width variations as an analog for flow {in fractures and horizontally heterogeneous aquifers}. Using phase-plane analysis, we obtain second-kind self-similar solutions to describe the evolution of the gravity current's shape during both the spreading (pre-closure) and leveling (post-closure) regimes. The self-similar theory is compared to numerical simulations of the partial differential equation governing the evolution of the current's shape (under the lubrication approximation) and to table-top experiments. Specifically, simulations of the governing partial differential equation from lubrication theory allow us to compute a pre-factor, which is \textit{a priori} arbitrary in the second-kind self-similar transformation, by estimating the time required for the current to enter the self-similar regime. With this pre-factor calculated, we show that theory, simulations and experiments agree well near the propagating front. In the leveling regime, the current's memory resets, and another self-similar behavior emerges after an adjustment time, which we estimate from simulations. Once again, with the pre-factor calculated, both simulations and experiments are shown to obey the predicted self-similar scalings. For both the pre- and post-closure regimes, we provide detailed asymptotic (analytical) characterization of the universal current profiles that arise as self-similarity of the second kind.\\ 
\end{abstract}

\maketitle


\section{Introduction}
\label{sec:intro}

\begin{figure}[t]
  \centerline{\includegraphics[width=0.9\textwidth]{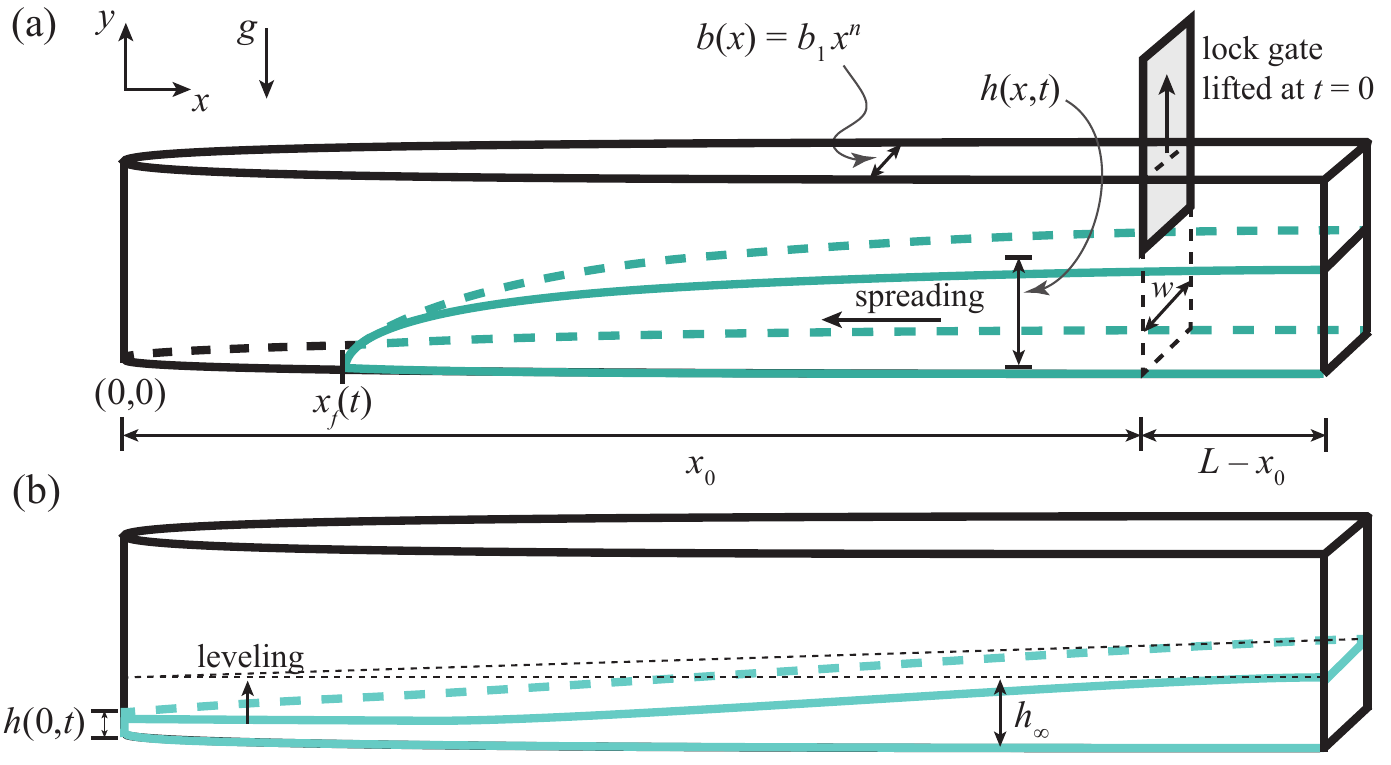}}
  \caption{Schematic of a slender horizontal channel (a Hele-Shaw cell) with varying gap thickness $b(x)$ in which a viscous gravity current is (a) spreading (pre-closure) and (b) leveling (post-closure). The shape of the current and the position of the current's moving front are denoted by $h(x,t)$ and $x_f(t)$, respectively. A Newtonian fluid is initially contained behind a lock gate at $x=x_0$. 
Upon release at $t=0^+$, it spreads by propagating toward the origin in the direction of decreasing gap thickness (i.e., the $-x$ direction). Upon reaching the origin at the closure time $t_c$, $x_f(t_c)=0$, and the current begins to level, with $h(0,t>t_c)>0$, and eventually reaches a flat asymptotic state, $h(x,t\to\infty)=h_\infty$. Gravity is directed in the $-y$ direction.}
\label{fig:channel}
\end{figure}

The gravity-driven spreading of viscous fluids has been of significant interest in the literature starting from the second half of the 20th century (see e.g., \cite{Simpson1999,Huppert2006,Woods2015}). Typically, one considers the horizontal spreading of a heavier fluid beneath a lighter one (i.e., there is a density difference $\Delta \rho > 0$ between the fluids). The motion of the denser fluid is dictated by a balance of buoyancy (gravity) and viscous forces at a low ``effective'' Reynolds number and a large Bond number. {For example, this kind of gravity current flow scenario, as illustrated in Fig.~\ref{fig:channel}, could model confined brine--CO\textsubscript{2} displacement in a heterogeneous aquifer \cite{Phillips1991,Woods2015}. Another application is flow in hydraulic fractures, for which the width $b(x) \sim (x-x_\mathrm{tip})^{1/2}$ as $x\to x_\mathrm{tip}$ \citep{GD2000}, where $x_\mathrm{tip}=0$ in Fig.~\ref{fig:channel}.} In these situations, the viscous gravity currents are characterized by slender geometric profiles (i.e., they have small aspect ratios such that $h/L \ll 1$, where $h$ and $L$ are typical vertical and horizontal length scales, respectively). Therefore, these flows can be modeled within the context of lubrication theory (see, e.g., \cite{ODB97,L07}). Generically, one obtains a nonlinear, parabolic partial differential equation (PDE) for the gravity current's shape $h$  (i.e., the vertical extent of the denser fluid, as illustrated in Fig.~\ref{fig:channel}) as a function of the flow-wise coordinate $x$ and time $t$. 

Being governed by a parabolic (irreversible) PDE viscous gravity currents `forget' the initial conditions from which they evolve, for some intermediate range of $t$. This observation is the concept of \emph{intermediate asymptotics} \cite{Barenblatt1972,Barenblatt1979}. It follows that a \emph{self-similar} (`universal', since it is independent of initial conditions) current profile exists during this intermediate asymptotic time period. This profile is usually found by reducing the governing PDE to an ordinary differential equation (ODE) by a self-similarity transformation. If the similarity variable can be obtained by a scaling (dimensional) analysis, this kind of solution is known as a self-similar solution of the \emph{first} kind \cite[Ch.~3]{Barenblatt1979}. 

Self-similarity of the first kind has been used to analyze the propagation of viscous gravity currents in a variety of physical scenarios. For example, a wealth of exact and approximate self-similar solutions for Newtonian gravity currents in porous media are available in Refs.~\cite{Barenblatt1952,Huppert1995,Anderson2003,Lyle2005,Vella2006,Hesse2007,Anderson2010,DeLoubens2011,Ciriello2013,Zheng2013,zcs14,zfccs15,Furtak-Cole2018,McCue2019,Sahu2020}, amongst many others. If the similarity variable cannot be obtained by a scaling analysis, then the problem represents self-similarity of the \emph{second} kind \cite[Ch.~4]{Barenblatt1979}. \citet{Gratton1990} classified a variety of second-kind self-similar solutions for viscous gravity currents by transforming the governing nonlinear parabolic PDE in such a way as to yield an autonomous nonlinear ODE in the plane. More recently, \citet{zcs14} extended Gratton's approach to gravity-driven spreading in porous media that exhibit permeability and porosity variations in the flow-wise direction, via the analogy between Darcy and Hele-Shaw flow in two-dimensions \cite{Bear1972,Homsy1987}. These types of flows require the identification of a length scale, related to the distance the gravity current must travel to reach the dead-end of the flow geometry ({e.g., $L-x_0$ in} Fig.~\ref{fig:channel}). Hence, a complete self-similar solution of the first kind cannot be obtained by scaling arguments, and some `memory' of the initial conditions is retained (via the dependence upon, e.g., $L-x_0$). Yet, it was shown (via theory, simulations and experiments) in Ref.~\cite{zcs14} that these gravity currents nevertheless do enter a self-similar propagation regime. Permeability of the substrate can also be taken in to account, yielding further second-kind self-similar solutions \cite{zss2015}. Most recently, the phase-plane formalism has been applied to converging gravity currents of non-Newtonian (power-law) fluids, showing agreement between second-kind self-similarity theory and experiments \citep{Longo2021}.

In this work, we study self-similarity of the second kind for gravity-driven viscous flows in Hele-Shaw cells with flow-wise shape (width) variation, as illustrated in Fig.~\ref{fig:channel}. We provide a detailed discussion on the self-similar dynamics of the post-closure (leveling) process, which occurs after the current reaches the dead-end of the Hele-Shaw cell (or, using the axisymmetric case's terminology, the `hole' is completely closed). {Although self-similar profiles for a converging gravity current were numerically calculated by \citet{Gratton1990,zcs14,Longo2021} by integrating the phase plane ODE, the mathematical structure of these solutions and a fundamental understanding of which experimentally-measured quantities one should correlate in post-closure (and why) is lacking. When successful, previous studies relied on analogy and intuition to understand the post-closure self-similar behavior. To fill this knowledge gap, this work provides a complete mathematical analysis of the problem, including analytical results obtained using dynamical systems theory, to justify \emph{both} the anomalous exponents \emph{and} the undetermined pre-factors of the similarity variables.} 

The model system is reviewed in Section~\ref{sec:model}. Then, in Section~\ref{sec:shape_calc}, asymptotic analytical expressions are derived for the profile shapes, both in the pre- and post-closure regimes. These solutions are compared with table-top experimental measurements and numerical solutions of the governing lubrication PDE in Section~\ref{sec:updates}. We find that, after an initial transition period, the shape of the current indeed approaches a regime in which the predicted scalings of the second-kind self-similarity are demonstrated, for both the spreading and leveling processes. Importantly, although the second-kind self-similar transformation contains an \textit{a priori} arbitrary pre-factor, simulations of the governing lubrication PDE allow us to compute this pre-factor (in Section~\ref{sec:SSnum}) in terms of a time scale characterizing the transition process towards the intermediate asymptotics (and the various problem parameters).

\clearpage

\section{Mathematical model for second-kind self-similarity of converging gravity currents}
\label{sec:model}

\subsection{Preliminaries: Notation and terminology}
\label{sec:prelim}

Consider the spreading of a viscous fluid in a Hele-Shaw (HS) cell of a varying width in the $x$-direction as in \cite{zcs14}, illustrated in Fig.~\ref{fig:channel}. The width $b$ of the HS cell varies as a power-law in $x$, i.e., $b(x) = b_1 x^n$ with $b_1>0$ and $n \in (0,1)$ being constant. The HS cell is vertically unconfined and possesses an impermeable bottom surface. The fluid spreading in the HS cell is assumed to be Newtonian (constant viscosity $\mu$). Our model below relies on the lubrication approximation, which requires that the reduced Reynolds numbers $(h/L)^2Re, (w/L)^2Re \ll 1$, where the current's (slender) flow-wise and transverse aspect ratios are $h/L, w/L \ll 1$, and $w$ is a transverse characteristic dimension. {Surface tension effects have been neglected under the assumption of a large Bond number $Bo\gg1$, which indicates the dominance of gravity-driven spreading.} By satisfying these conditions, we may use a gap-averaged formulation as is standard for Hele-Shaw problems. {Furthermore, we neglect the drag from the bottom plate by considering $h$ to be sufficiently large compared to $b$, so that the sidewall drag is dominant, and the hydraulic conductivity is set by $\Delta \rho g b^2/(12\mu)$ \cite{Bear1972,zcs14}.}

The fluid is assumed to be initially contained behind a release (or, lock gate) located at $x=x_0$, as depicted in Fig.~\ref{fig:channel}. Upon opening the gate, the current slumps and spreads `leftwards', towards the origin at $x = 0$, where the width vanishes ($b(0) = 0$). The total fluid volume $\mathcal{V}$ remains constant within the domain $x\in[0,L]$. The moving front of the current is initially at $x_f(0) = x_0$. The current spreads until it reaches the origin at a time of \emph{closure}, $t=t_c$ (also termed `touch-down' time in \cite{zcs14}); this period $t\in[0,t_c)$ is hereafter refereed to as \emph{pre-closure}. Thereafter, the current levels at $x=0$ until $h(x,t \to \infty) = const.$; this period $t\in[t_c, \infty)$ is hereafter referred to as \emph{post-closure}. 

The closure time $t_c$ can be determined numerically from simulations, or it can be obtained from experiments \cite{Diez1992,zcs14}. This additional time scale can be \emph{infinite}, if the current never reaches the origin. For the present purposes, it is assumed that $t_c<\infty$ in the absence of capillary effects or substrate drainage. Regardless, $t_c$ (or, equivalently, $L-x_0$) emerges as an extra time (or, equivalently, length) scale, hence the scaling analysis (which might be used to seek a self-similar solution) becomes ambiguous. For example, $x_0$ now appears in the global mass conservation constraint $\int_0^L h(x,t)b(x)\,\mathrm{d}x = \int_{x_0}^L h(x,0)b(x)\,\mathrm{d}x=\mathcal{V}=const$. Indeed, \emph{complete} self-similarity with respect to a single similarity variable cannot exist in such a problem \cite{Barenblatt1979}. Nevertheless, self-similar behavior can be expected and has been observed \cite{zcs14}. The phase-plane formalism can be used to explain the observed self-similarity; see, e.g., Gratton and Minotti's \cite{Gratton1990} application of this method, which is lucidly explained in the book by Sedov \cite{Sedov1993}. Moreover, as we shall now show, the self-similar analysis of the governing equations in an appropriate phase-plane can predict the existence of \emph{two} self-similar regimes, one in pre- and one in post-closure. It is expected that any self-similarity variable of the second kind explicitly features $t_c$ .

\subsection{Similarity transformation and phase-plane analysis}
\label{sec:phase_plane_deriv}

Under the lubrication approximation, a width-averaged model can be obtained (see, e.g., \cite[Ch.~6]{L07}). To apply the phase-plane formalism \cite{Gratton1990}, it is convenient to follow \cite{zcs14} (see also \cite{Diez1992}) and start with the formulation of the model as a system of two first-order equations. The pressure distribution is hydrostatic, so the fluid flux and the continuity equation, respectively, take the form:%
\begin{subequations}\begin{align}
  &u = - \frac{\Delta \rho g b_1^2}{12 \mu} x^{2n} \frac{\partial h}{\partial x} , \label{eq:udhdx} \\[3pt]
  &\frac{\partial h}{\partial t} + \frac{1}{x^n} \frac{\partial}{\partial x}(x^nhu) = 0.
 \label{eq:conserv}
\end{align}\label{eq:pdes}\end{subequations}
Here, $u=u(x,t)$ is the width-averaged (also termed `superficial') velocity. This model is essentially `one-phase', as the dynamics of the upper fluid (air in the HS cell in Fig.~\ref{fig:channel}) is neglected. {At this stage, we do not state any boundary conditions for Eqs.~\eqref{eq:udhdx} and \eqref{eq:conserv}, as the upcoming self-similarity analysis is \emph{local}, applying near the front of the current.}

To study both spreading and leveling of the current, we assume that $t_c < \infty$, and introduce a shifted time $\tau = t_c - t$. This shifted time $\tau$ represents the time remaining until the current reaches the origin, and its definition necessitates the use of numerical simulations (or experiments) to determine the closure time $t_c$ \emph{a posteriori}. The next step in the phase-plane formalism is to render Eqs.~\eqref{eq:udhdx} and \eqref{eq:conserv} dimensionless by using the \emph{independent variables as dimensional scales}. Specifically, we introduce the transformations
\begin{subequations}
\begin{align}
	u(x,t) &= \frac{x}{\tau} U(x,\tau), \label{eq:u_U_transform}\\
	h(x,t) &= \left(\frac{12 \mu}{\Delta \rho g b_1^2}\right)\frac{x^{2(1-n)}}{\tau} H(x,\tau), \label{eq:h_H_transform}
\end{align}\label{eq:dimless_var_1}\end{subequations}
where $U(x,\tau)$ and $H(x,\tau)$ are the dimensionless analogs to the width-averaged velocity and the current height, respectively. Note that since, $u \leq 0$ for the current spreading towards the origin, $U \leq 0$ as well, while $H \geq 0$ for the equal and opposite reason.

Substituting Eqs.~\eqref{eq:u_U_transform} and \eqref{eq:h_H_transform} into Eqs.~\eqref{eq:udhdx} and \eqref{eq:conserv}, the governing equations are re-written in terms of $H$ and $U$ (see Ref.~\cite{zcs14}):
\begin{subequations}\begin{align}
    x \frac{\partial H}{\partial x} + 2 (1-n) H + U &= 0, \\[3pt]
    \tau \frac{\partial H}{\partial \tau} - H - x \frac{\partial}{\partial x}(H U) - (3- n ) HU &= 0.
\end{align}\label{eq:ss_prelim_eqns}\end{subequations}
Anticipating a self-similar solution, a \emph{second-kind} self-similarity variable of the form $\xi = x/\tau^\delta$, such that $\xi>0$, is introduced; at the moving front $\xi = \xi_f \equiv x_f(t)/\tau^\delta$. Critically, $\delta$ is \emph{unknown} here, and $\xi$ explicitly features the scale $t_c$ through $\tau$. The assumption of self-similarity now necessitates that $H = H(\xi)$ and $U = U(\xi)$. This assumption allows the governing Eqs.~\eqref{eq:ss_prelim_eqns} to be reduced to a system of one-way coupled ODEs:
\begin{subequations}\begin{align}
  \frac{\mathrm{d}U}{\mathrm{d}H} &= \frac{H[(n+1)U - 2(1-n)\delta + 1] - U(U + \delta)}{H[2(1-n)H+U]} , \label{eq:dUdH} \\[3pt]
  \frac{\mathrm{d} \ln \xi }{\mathrm{d}H} &= - \frac{1}{U+2(1-n)H} .
\label{eq:dlnxidH}
\end{align}\label{eq:ppode}\end{subequations}
Equation~\eqref{eq:dUdH} represents an autonomous ODE for $U(H)$ depending on a parameter $\delta$. Once $U(H)$ is known, Eq.~\eqref{eq:dlnxidH} is used to find $\xi(H)$, from which the self-similar profiles $H(\xi)$ and $U(\xi)$ can be reconstituted from $U(H)$ (i.e., `re-parameterized' in terms of $\xi$). However, before any of these ODEs can be solved, a suitable set of boundary conditions (BCs) must be specified. Note that the success of the self-similar transformation, which we used to arrive at the system~\eqref{eq:ppode},  already suggests that a self-similar solution might exist. However, since $\delta$ is unknown, the problem becomes an \emph{eigenvalue problem}. With a suitable set of BCs, it is expected that both $U(H)$ and $\delta$ emerge as an `eigenpair' solution to Eq.~\eqref{eq:dUdH} through a global bifurcation \cite{Kevrekid1987}.


\subsection{Critical points in the phase-plane and their physical interpretation}
\label{sec:critical_points}

Since Eq.~\eqref{eq:dUdH} is a planar ODE, it follows that BCs arise as beginning and end points of integral curves in the $(H,U)$ plane. In order to identify the integral curves of physical significance in this $(H, U)$ plane, i.e., those solutions $U(H)$ that correspond to an observable self-similar behavior, the ODE's critical points in the phase plane must be found. Following \cite{Gratton1990,zcs14}, critical points are found by requiring that the numerator and denominator in Eq.~\eqref{eq:dUdH} vanish simultaneously. Thus, the points denoted below as O, A and B are obtained. The final critical point D is obtained by letting the denominator in Eq.~\eqref{eq:dUdH} go to $\infty$. In summary:%
\begin{subequations}\begin{align}
\mathrm{O}\; :\; (H,U) &= (0,0),\label{eq:pt_O}\\[3pt]
\mathrm{A}\; :\; (H,U) &= (0,-\delta),\label{eq:pt_A}\\[3pt]
\mathrm{B}\; :\; (H,U) &= \left(\frac{1}{2(1-n)(3-n)}\,,\, -\frac{1}{3-n}\right),\label{eq:pt_B}\\[3pt]
\mathrm{D}\; :\; (H,U) &= \left( -\infty,\frac{2(1-n)\delta - 1}{n+1} \right).\label{eq:pt_D}
\end{align}\label{eq:critical_pts}\end{subequations}
Points A and D depend upon the eigenvalue $\delta$, showing how the BCs will `conspire' with the ODE to determine the appropriate eigenpair solution.

As described in \cite{zcs14}, Point O corresponds to the instant of time at which the current reaches the point of `closure' at the channel's origin (corresponding to $x = 0$ or $\xi = 0$). Meanwhile, Point A corresponds to the moving front of a spreading current (at $x=x_f(t)$ or $\xi=\xi_f$). Point B does not have a physical interpretation in the present context. Point D corresponds to the leveling (post-closure) behavior. The integral curves connecting O and A, and D and O in the phase plane thus represent the sought self-similar solutions to the problem during pre-closure ($t<t_c$) and post-closure ($t>t_c$), respectively.

\begin{figure}[t]
\centerline{\includegraphics[width=0.75\textwidth]{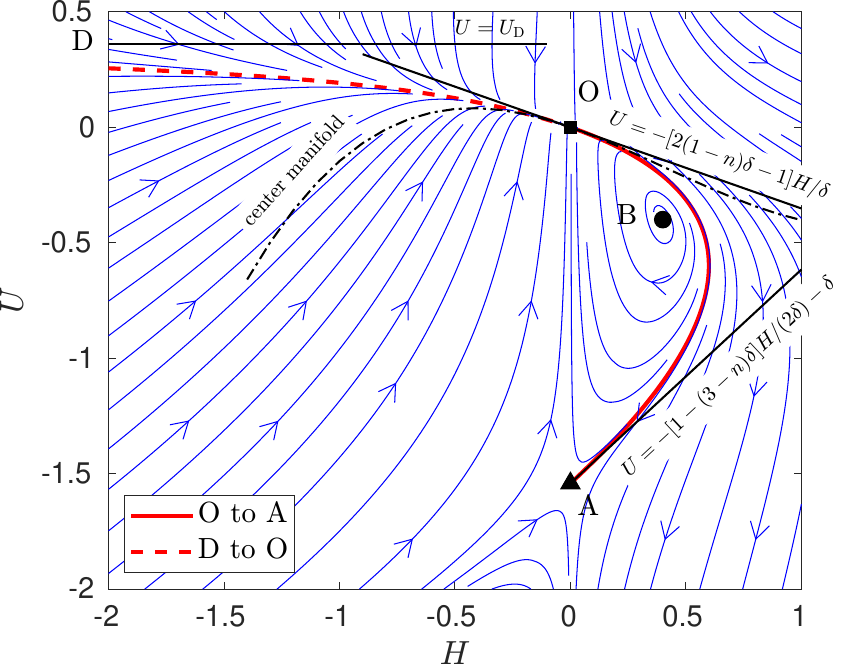}}
\caption{Phase-plane portrait of the ODE~\eqref{eq:dUdH} describing the self-similar propagation of a Newtonian current in a shaped HS cell with width exponent $n = 0.5$. The eigenvalue $\delta$ appearing in the definition of the similarity variable $\xi$ was calculated to be $\approx 1.542269$. The integral curves O to A (solid red) and D to O (dashed red) represent self-similar pre- and post-closure solutions, respectively. The dash-dotted curve labeled `center manifold' is the higher-order approximation near Point O derived in Appendix~\ref{app:pointO}. Solid black lines, correspond to the various linear approximations to the integral curves near Points A, O, and D as described in the text.}
\label{fig:pplane}
\end{figure}

Having identified the integral curves of interest, the task of finding a self-similar solution has been reduced to a nonlinear eigenvalue problem. Specifically, the question now is, given $n$, what value(s) of $\delta$ allow for the existence of phase-plane curves that connect Point O to A and Point D to O? The nonlinear eigenvalue problem can be solved using a `shooting' procedure (for details see, \cite[Section 2.1.2]{zcs14} or \cite{Gratton1990,Kevrekid1987}). For instance, for the case of $n=0.5$, we find that $\delta \simeq 1.542269$, to single precision. The corresponding phase-plane is depicted in Fig.~\ref{fig:pplane}, highlighting the sought-after phase-plane solutions, which were computed numerically as described in, e.g., \cite{zcs14}. Thus, the existence of two distinct self-similar regimes has been predicted, and the value of the similarity exponent $\delta$ has been determined.

%

\section{Calculation of the pre- and post-closure self-similar solutions}
\label{sec:shape_calc}

In this section, we calculate the \emph{shape} of the second-kind self-similar gravity current, during both pre- and post-closure, starting from the asymptotics near critical points of the phase-plane ODE. These shapes have not been discussed in the literature \cite{zcs14} for the variable-width HS cell.
 
To study the asymptotic behavior near the ODE's critical points in Eq.~\eqref{eq:critical_pts}, and thus obtain approximate analytical expressions for the integral curves $U(H)$ from which the self-similar gravity current shape follows, it is convenient to rewrite Eq.~\eqref{eq:dUdH} as an autonomous system:
\begin{equation}
\frac{\mathrm{d}}{\mathrm{d}s}\begin{pmatrix} H\\ U \end{pmatrix} 
= \begin{pmatrix} H[2(1-n)H + U]\\ H[(n+1)U - 2(1-n)\delta + 1] - U(U+\delta) \end{pmatrix},
\label{eq:HU_system}
\end{equation}
where $s$ is a `dummy' parametric variable that can be understood as being a monotonic function of time (or distance) along an integral curve in the phase plane. 
Linearizing the system in Eq.~\eqref{eq:HU_system}, we find the Jacobian $\mathbb{J}$, evaluated about some generic point $(H^*,U^*)$, to be
\begin{equation}
\mathbb{J}(H^*,U^*)
= \begin{pmatrix}
 4 H^* (1-n)+U^* & H^* \\
 (n+1)U^* - 2(1-n) \delta + 1 & H^* (n+1) - 2U^* - \delta
\end{pmatrix}.
\label{eq:jacobian}
\end{equation}

Next, in Section~\ref{sec:pointA}, we turn to the asymptotic behavior near Point A, which provides the shape of a spreading gravity current near the moving front (pre-closure). Then, in Section~\ref{sec:pointD}, we study the asymptotic behavior near Point D,  which provides the shape of a leveling gravity current after the front reaches the origin (post-closure). For completeness, in Appendix~\ref{app:pointO}, we provide the  asymptotics (including a center manifold reduction) at Point O.

\subsection{Point A: Spreading behavior}
\label{sec:pointA}

In this subsection, we study the asymptotic behavior near Point A, as defined in Eq.~\eqref{eq:pt_A}, of the integral curve connecting Point O to Point A, which corresponds to the self-similar shape of the spreading gravity current, before the front has reached the origin \cite{zcs14,Gratton1990}. The shape of the current near the moving front is determined by the asymptotic behavior of the heteroclinic trajectory in the $(H,U)$ plane near Point A. 

To determine the asymptotics, we linearize the system in Eq.~\eqref{eq:HU_system} near Point A, which is given in Eq.~\eqref{eq:pt_A} as $(H^*,U^*)=(0,-\delta)$. The Jacobian from Eq.~\eqref{eq:jacobian} becomes
\begin{equation}
\mathbb{J}(0,-\delta) = \begin{pmatrix}
 -\delta & 0 \\
 1-(3-n)\delta & \delta
\end{pmatrix}.
\label{eq:jacobian_A}
\end{equation}
The eigenvalues of $\mathbb{J}(0,-\delta)$ are
\begin{equation}
	\lambda_{1,2} = \pm \delta,
\end{equation}
while the corresponding eigenvectors are 
\begin{equation}
	\bm{v}_1 = \pm \begin{pmatrix} 0\\ 1 \end{pmatrix},\qquad \bm{v}_2 = \pm \begin{pmatrix} \displaystyle\frac{2 \delta }{(3-n)\delta-1}\\ 1 \end{pmatrix}.
\label{eq:evec_ptA}
\end{equation}
Clearly, if any eigenvector at Point A is to correspond to the direction of the incoming heteroclinic trajectory, it would be $\bm{v}_2$ with the ``$+$'' sign because $\lambda_2 = -\delta < 0$ indicates this is a \emph{stable} direction \emph{into} Point A.

Now, from $\bm{v}_2$ in Eq.~\eqref{eq:evec_ptA}, the stable manifold at Point A has the slope
\begin{equation}
\frac{\mathrm{d}U}{\mathrm{d}H} \sim - \frac{1 - (3-n)\delta}{2 \delta } \quad\Rightarrow\quad U \sim - \left[\frac{1 - (3-n)\delta}{2 \delta }\right] H -\delta\quad\text{as}\quad (H,U) \to (0,-\delta),
\label{eq:slope_OA}
\end{equation}
where the constant of integration was set by requiring that the trajectory go through Point A. 
In passing, we note that Eq.~\eqref{eq:slope_OA} is consistent with \cite[Eq.~(2.17)]{zcs14} (i.e., $U \sim H - \delta$), because the latter is the limit of the former as $n\to 1^-$ with $\delta \sim 1/[2(1-n)]$.

Then, using Eqs.~\eqref{eq:dimless_var_1} and \eqref{eq:udhdx}, we rewrite Eq.~\eqref{eq:slope_OA} as
\begin{equation}
-\frac{\tau}{x}\frac{\Delta\rho g b_1^2}{12\mu} x^{2n} \frac{\partial h}{\partial x} \sim -\left[\frac{1 - (3-n)\delta}{2 \delta }\right]\left(\frac{\Delta\rho g b_1^2}{12\mu}\right)\frac{\tau}{x^{2(1-n)}} h(x,t) - \delta .
\label{eq:asymp_h_ptA}
\end{equation}
Restricting to the behavior as $x\to x_f^{+}$, we obtain, form Eq.~\eqref{eq:asymp_h_ptA}, a first-order ODE in $x$ for $h(x,t)$:
\begin{equation}
\frac{\partial h}{\partial x} - \frac{1}{x}\left[\frac{1 - (3-n)\delta}{2 \delta }\right] h =  \frac{1}{x}\left(\frac{12\mu}{\Delta\rho g b_1^2}\right)\frac{x^{2(1-n)}}{\tau}\delta,\qquad\text{such that}\qquad h(x_f,t) = 0.
\label{eq:h_ode_pc}
\end{equation}
Then, using an integrating factor, we find the solution to the ODE~\eqref{eq:h_ode_pc}, which provides the asymptotic shape of the spreading current near the moving front:
\begin{equation}
h(x,t) \sim \left[\left(\frac{12\mu}{\Delta\rho g b_1^2}\right)\frac{x^{2(1-n)}}{\tau}\right] \underbrace{\frac{2 \delta ^2}{(7- 5n)\delta - 1}
\left[ 1 -  (x/x_f)^{[(5n-7)\delta+1]/(2\delta)}\right]}_{H_\mathrm{front}(\xi/\xi_f)}
\quad\text{as}\quad x\to x_f^{+},
\label{eq:hxt_tip}
\end{equation}
which is valid for $x\ge x_f > 0$, and the current propagates right to left (as shown in Fig.~\ref{fig:channel}).
Additionally, from Eqs.~\eqref{eq:udhdx} and \eqref{eq:hxt_tip}, we find the width-averaged velocity near the moving front:
\begin{equation}
u(x,t) \sim \frac{x}{\tau} 
 \underbrace{\frac{2 \delta ^2}{(7- 5n)\delta - 1}
 \left\{ 2(n-1) + \left[\frac{1 - (3-n)\delta}{2 \delta}\right] (x/x_f)^{[(5n-7)\delta+1]/(2\delta)} \right\}}_{U_\mathrm{front}(\xi/\xi_f)} \quad\text{as}\quad x\to x_f^{+}.
 \label{eq:uxt_tip}
\end{equation}

In Eqs.~\eqref{eq:hxt_tip} and \eqref{eq:uxt_tip}, we have denoted by $H_\mathrm{front}(\xi/\xi_f)$ and $U_\mathrm{front}(\xi/\xi_f)$ the approximate analytical expressions, near the current front ($\xi\to\xi_f$), for the self-similar solutions of the second kind. 


\subsection{Point D: Leveling behavior}
\label{sec:pointD}

In this subsection, we study the asymptotic behavior near Point D, as defined in Eq.~\eqref{eq:pt_D}, of the integral curve from Point D to Point O in the $(H,U)$ plane, which corresponds to the self-similar shape of the leveling gravity current, after the front has reached the origin \cite{Diez1992,Diez1992b}. We first note that, to analyze the leveling (post-closure) self-similar behavior, we must define a new self-similarity variable $\zeta = x/(-\tau)^\delta$ since $\tau < 0$ for $t>t_c$. Note that our definition of $\zeta$ (like $\xi$) is non-negative, i.e., $\zeta\ge 0$, unlike prior literature \cite{Diez1992,Gratton2010}.

To determine the post-closure asymptotic behavior, we expand the right-hand side of Eq.~\eqref{eq:dUdH} near Point D:
\begin{equation}
\frac{\mathrm{d}U}{\mathrm{d}H} \sim \frac{-U_\mathrm{D}(U_\mathrm{D}+\delta)}{2(1-n)H^2}
\quad\Rightarrow\quad
U(H) \sim U_\mathrm{D}\left[1 + \frac{(U_\mathrm{D}+\delta)}{2(1-n)H} \right]\quad\text{as}\quad (H,U) \to (-\infty,U_\mathrm{D}),
\label{eq:U_near_D_2}
\end{equation}
where, for convenience, we have made the definition $U_\mathrm{D} \equiv [2(1-n)\delta - 1]/(n+1)$. This asymptotic approximation is not shown in Fig.~\ref{fig:pplane} because it is only valid for $H\to-\infty$, values far outside the plotting range.

Now, substituting $U(H)$ from Eq.~\eqref{eq:U_near_D_2} into Eq.~\eqref{eq:dlnxidH} and using a Taylor-series expansion, we obtain
\begin{equation}
\frac{\mathrm{d}\ln \zeta}{\mathrm{d}H} = - \frac{1}{U+2(1-n)H} \sim 
- \frac{1}{2(1-n)H} \left[ 1 - \frac{U_\mathrm{D}}{2(1-n)H}+\cdots\right],
\qquad H \to -\infty.
\end{equation}
We can integrate the latter ODE and write the shape explicitly using the principal branch (to set the $+$ and $-$ signs) of the Lambert-$W$ function \cite{Corless1996} as
\begin{equation}
H \sim -\frac{U_\mathrm{D}}{2(1-n)} \left\{W_0\left(\frac{U_\mathrm{D}}{2(1-n)}(\zeta/\zeta_f)^{2(1-n)}\right) \right\}^{-1},\qquad \zeta/\zeta_f \to 0^+.
\label{eq:H_ptD}
\end{equation}
Note that {$1/W_0(\varkappa) = 1/\varkappa + 1 + \mathcal{O}(\varkappa)$} as $\varkappa\to0$. Thus, we can also write $H\sim - (\zeta/\zeta_f)^{-2(1-n)}$ asymptotically as $\zeta/\zeta_f \to 0^+$, {if we keep just one term in the Taylor series of $1/W_0$}.

Next, following the earlier procedure (from Section~\ref{sec:pointA}), substituting Eqs.~\eqref{eq:U_near_D_2} and \eqref{eq:H_ptD} into Eqs.~\eqref{eq:dimless_var_1} and keeping only the leading-order terms, we obtain
\begin{subequations}\begin{align}
u(x,t) &\sim \frac{x}{\tau} \overbrace{U_\mathrm{D}\left[1 - \frac{(U_\mathrm{D}+\delta)}{U_\mathrm{D}} W_0\left(\frac{U_\mathrm{D}}{2(1-n)}(x/x_f)^{2(1-n)}\right) \right]}^{U_\mathrm{front}(\zeta/\zeta_f)}, \label{eq:u_ptD}\\
h(x,t) &\sim \left(\frac{12\mu}{\Delta\rho g b_1^2}\right)\frac{x^{2(1-n)}}{\tau} \underbrace{\left[\frac{-U_\mathrm{D}}{2(1-n)}\right] \left\{W_0\left(\frac{U_\mathrm{D}}{2(1-n)}(x/x_f)^{2(1-n)}\right) \right\}^{-1}}_{H_\mathrm{front}(\zeta/\zeta_f)}, \label{eq:hxt_tip_post}
\end{align}\label{eqs:post_asymp}\end{subequations}
where we recall that $\tau < 0$ in the post-closure regime, so $h > 0$ even though $H_\mathrm{front}<0$. 
Observe that Eq.~\eqref{eq:hxt_tip_post} predicts a slope $\partial h/\partial x \sim x^{1-2n}$ as $x\to0$, {which is obtained using the two-term expansion $1/W_0(\varkappa)\sim 1/\varkappa + 1$}. Therefore, $\partial h/\partial x$ is a finite constant as $x\to0$ for the special case $n=1/2$, while $\partial h/\partial x \to 0$ as $x\to0$ for $n<1/2$ and $\partial h/\partial x \to \infty$ as $x\to0$ for $n>1/2$. In all three cases, of course, the width-averaged velocity still vanishes as $x\to0$ (as required by the physical fact that this is the dead-end of the HS cell) due to the $x^{2n}$ term pre-multiplying $\partial h/\partial x$ in Eq.~\eqref{eq:udhdx}. 

\subsection{Computing the self-similar solutions}
\label{sec:ODE_solve}

In this subsection, to complement the asymptotics obtained in Sections~\ref{sec:pointA} and \ref{sec:pointD}, we solve Eqs.~\eqref{eq:dUdH} and \eqref{eq:dlnxidH} numerically to obtain the self-similar gravity current profile $H$, either pre- or post-closure, depending on the BCs applied (recall Fig.~\ref{fig:pplane}). It is inconvenient to go back and solve Eq.~\eqref{eq:dlnxidH} for $\xi(U)$, after solving Eq.~\eqref{eq:dUdH}, to re-parametrize $U(\xi)$ and $H(\xi)$. A mathematical `trick' can be used to avoid this inconvenience. First, the similarity variable $\xi$ is scaled by its value at the front, i.e., $\xi_f$. Now, the channel origin is defined as the point at which $\xi/\xi_f = 0$, while the moving front of the current is at $\xi/\xi_f = 1$ \footnote{Note that, in the post-closure regime, $\xi/\xi_f$ is replaced by $\zeta/\zeta_f$ in Eq.~\eqref{eq:HU_system3}, under our notation convention, but this notation change does not affect the ODE or its solution.}. Second, following e.g.~\cite{Slim2004}, we rewrite Eqs.~\eqref{eq:ppode} as 
\begin{equation}
\frac{\mathrm{d}}{\mathrm{d}\ln (\xi/\xi_f)}\begin{pmatrix} H\\ U \end{pmatrix} = \begin{pmatrix} -[2(1-n)H + U]\\ \big\{ -H[(n+1)U - 2(1-n)\delta + 1] + U(U+\delta) \big\}/H \end{pmatrix},
\label{eq:HU_system3}
\end{equation}
where $\delta$ is already known from having solved the nonlinear eigenvalue problem, as described in Section~\ref{sec:critical_points}.

\begin{figure}
\subfloat[][pre-closure]{\includegraphics[height=0.375\textwidth]{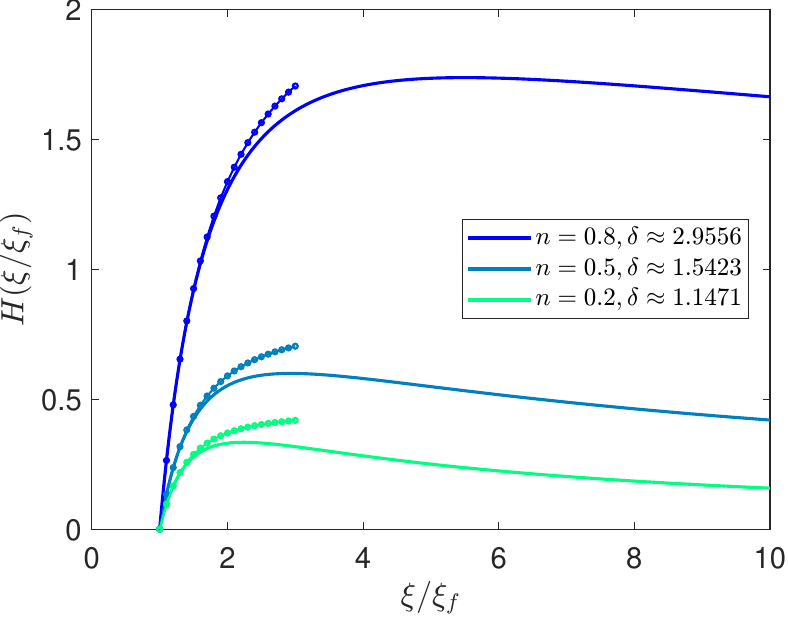}\label{fig:pre_n}}
\hfill
\subfloat[][post-closure]{\includegraphics[height=0.37\textwidth]{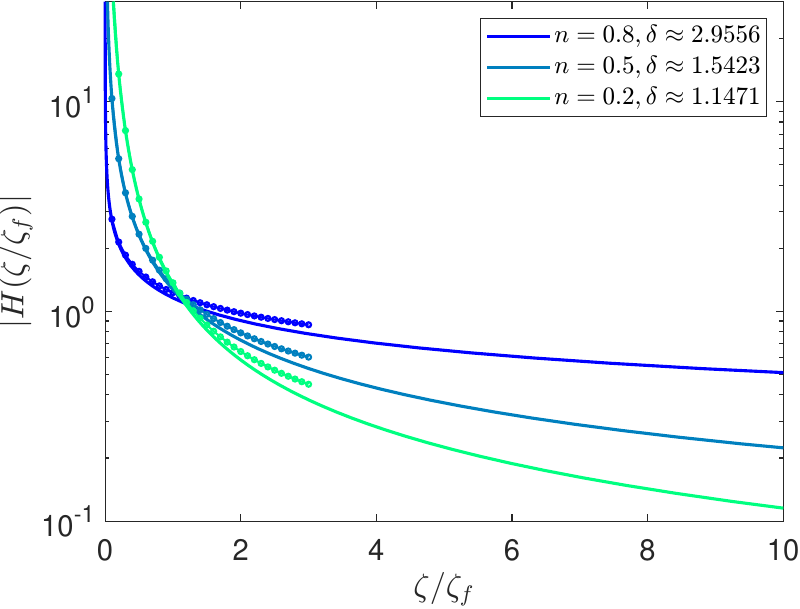}\label{fig:post_n}}
\caption{Impact of the HS cell geometry exponent $n$ on the self-similar profile $H$ (solid curves) obtained by numerically integrating Eq.~\eqref{eq:HU_system3}, for the (a) pre-closure and (b) post-closure regimes. Symbols correspond to the front asymptotics $H_\mathrm{front}$ from Eq.~\eqref{eq:hxt_tip} for (a) and from Eq.~\eqref{eq:hxt_tip_post} for (b). The self-similarity exponent $\delta$ was obtained, for each $n$, by an independent phase-plane analysis, as described in Section~\ref{sec:critical_points}. Note that the vertical scale in (b) is logarithmic to better highlight the variation of the shapes.}
\label{fig:multiple_n}
\end{figure}

For pre-closure, Eq.~\eqref{eq:HU_system3} is integrated `forward' from $\xi/\xi_f = 1 + \varepsilon$, where $\varepsilon$ is taken to be machine precision $\approx 10^{-16}$. For post-closure, the integration starts at $\zeta/\zeta_f = \varepsilon$. In both cases, the `initial' conditions for the integration are taken from the phase-plane asymptotics in Sections~\ref{sec:pointA} and \ref{sec:pointD}; specifically, Eq.~\eqref{eq:slope_OA} with $H(1 + \varepsilon) = \varepsilon$ for pre-closure, while $U(\varepsilon) = U_\mathrm{D}$ and $H(\varepsilon)=-1/\varepsilon^{2(1-n)}$ for post-closure from Eqs.~\eqref{eqs:post_asymp}. \textsc{Matlab}'s stiff ODE integration algorithm \texttt{ode23s} is employed. \textsc{Matlab}'s ODE solvers implement adaptive step control with relative and absolute tolerances \cite{Shampine97}, which we both set to $10^{-12}$, to ensure an accurate solution. Example (a) pre- and (b) post-closure numerical solutions for the self-similar profile $H$ are shown in Fig.~\ref{fig:multiple_n} for several representative values of the HS cell shape exponent $n$. The range of validity of the analytical expressions based on the front asymptotics is also highlighted. 

\section{Comparison between numerical and experimental results}
\label{sec:updates}

At this point in the analysis, we have found the two second-kind self-similar solutions that manifest in our problem. However, any additional analysis and interpretation of these solutions requires rescaling back to the physical variables, which requires the explicit knowledge of, at least, $t_c$ and $\xi_f$ (or $\zeta_f$). Therefore, the goal of this section is to determine these quantities.

First, in Section~\ref{sec:expts}, we describe the experiments to which the simulations in Section~\ref{sec:SSnum} are matched. The numerical simulations are based on the scalar lubrication PDE for $h(x,t)$ that results from eliminating $u$ between Eqs.~\eqref{eq:udhdx} and \eqref{eq:conserv}.  Comparisons of theory to experiments and simulations allow us to ascertain the physical validity of the second-kind self-similar solutions obtained from the phase-plane analysis in Section~\ref{sec:shape_calc}. In addition, the influence of the initial condition on the transition process towards self-similarirty will be illustrated by the PDE simulations.

\subsection{Experimental study}
\label{sec:expts}

\begin{figure}[ht]
\centerline{\includegraphics[width=0.75\textwidth]{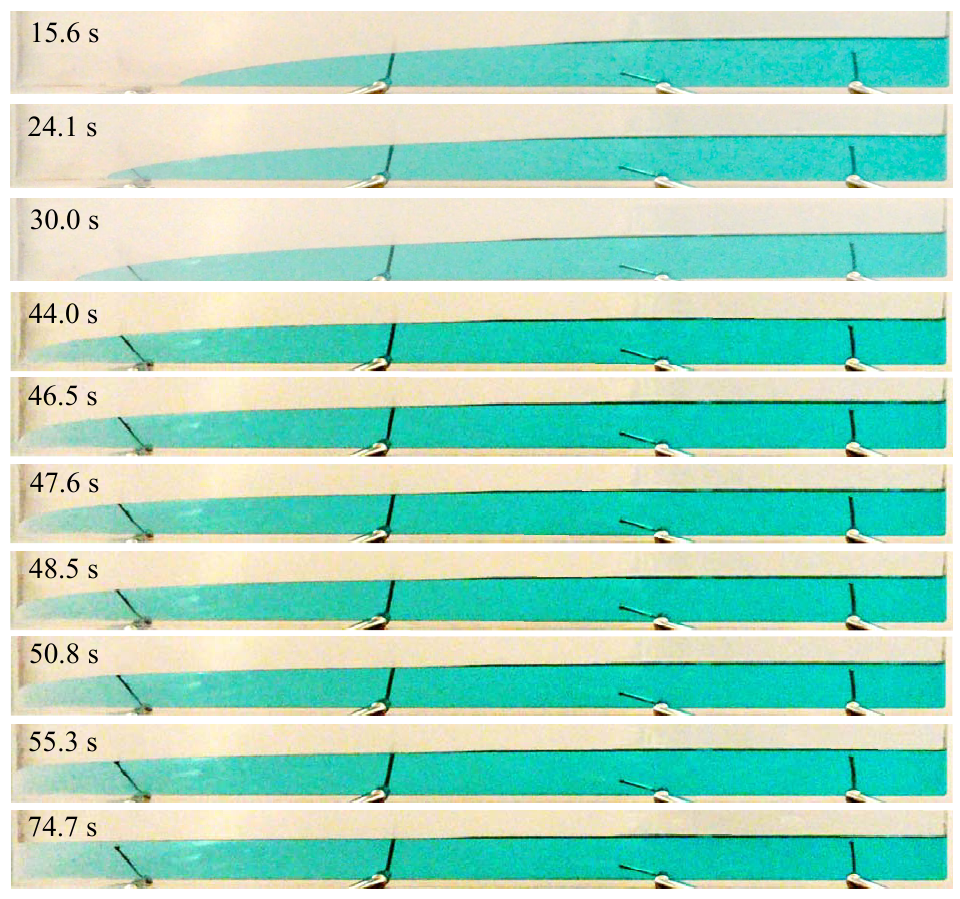}}
\caption{Experimental time-lapse (instant of time indicated on top left of each row) of a gravity current profile $h(x,t)$ spreading and leveling in a variable-gap Hele-Shaw cell with width variation exponent $n=0.5$. Glycerol was released into the cell and photographed during the pre- and post-closure. The flow is from right to left. From this experiment, a closure time of $t_c \approx 44.03$ s was determined. Therefore, the first three profiles show the spreading process (pre-closure, $t<t_c$), while the remaining ones represent the leveling process (post-closure, $t>t_c$).}
\label{fig:exptraw}
\end{figure}

An experiment was performed in the variable-width ($b(x)=b_1x^n$) horizontal HS cell depicted in Fig.~\ref{fig:channel}. A summary of the geometrical parameters and fluid properties necessary to analyze the experiment are given in Table~\ref{tb:expt}. Specifically, a constant fluid volume of $\mathcal{V}\approx 250$ cm\textsuperscript{3} was released from behind a lock gate and allowed to spread towards the origin (dead-end of the HS cell). The lock gate's location $x_0$ was taken to be comparable to the length of the cell $L$ to provide a longer distance over which the current's spreading behavior could be observed.  The fluid was colored using a food dye for digital post-processing. The shape of the current was photographed using a USB camera, yielding the profiles shown in Fig.~\ref{fig:exptraw}. The experiment was conducted with 100\% glycerol at 20\textdegree C, and the corresponding fluid properties were determined using standard fits \cite{Cheng_vis_2008,VolkAndreas2018,Takamura2012}.  

\begin{table}
\caption{Summary of the experimental parameters. The Hele-Shaw cell's dimensions and variation are specified through these values. The working fluid used was 100\% glycerol at 20\textdegree C.}
\centering
\begin{tabular}{p{6cm}p{3cm}p{3cm}}
\hline
Quantity & Value & Units \\ \hline
Channel length $L$               & 0.75     & m \\ 
Lock gate location $x_0$         & 0.4897   & m \\ 
Width coefficient $b_1$          & 0.01732  & m$^{1-n}$\\ 
Width exponent $n$               & 0.5      &-- \\
Total released mass              & 0.3155   &kg \\ 
Density difference $\Delta\rho$  & 1261     &kg/m\textsuperscript{3}\\
Dynamic viscosity $\mu$          & 1.412    &Pa$\cdot$s \\ 
Surface tension $\gamma$		 & $6.34 \times 10^{-6}$ &N/m \\
\hline\\
\end{tabular}
\label{tb:expt}
\end{table}

As discussed above, it is expected that, over the period of time that the current spreads, it evolved from its initial condition (attained as it sat behind the lock gate for $t\le0$) and enters the pre-closure self-similar regime as $t \to t_c$ ($\tau \to 0^+$). This notion and how it relates to the concept of self-similarity is discussed in more detail in Section~\ref{sec:SSnum}. Obtaining accurate experimental measurements of the current as it approaches the origin becomes challenging as the flow is accelerated by the converging nature (decreasing transverse width) of the HS cell. Nevertheless, experiments suggest that the current's closure time (i.e., the time it takes to the reach $x_f(t_c)=0$ from $x_f(0)=x_0$) is $t_c \approx 44.03$ s. Thereafter, the current enters the leveling, or post-closure, regime.

Before we can apply the theory based on Eqs.~\eqref{eq:pdes} and \eqref{eq:ppode}, it is critical to ensure that the experimental conditions fall within the assumptions of the lubrication approximation. As mentioned in Section~\ref{sec:prelim}, this requires that the aspect ratios and the reduced Reynolds numbers are small in both the vertical and transverse directions. To this end, consider the streamwise length scale to be $L \sim x_0$, the transverse length scale to be $w \sim b_1x_0^n$, and the vertical length scale to be the steady-state height of the fully leveled current $h \sim h_{\infty}$ (see Eq.~\eqref{eq:hinf} below and the attendant discussion). The velocity scale is then simply taken to be $x_0/t_c$, so that $Re = \Delta \rho\, x_0^2/(\mu t_c)$  \cite{zcs14}. For the experiments conducted in accordance with the parameters in Table~\ref{tb:expt}, $w/x_0 = 0.0247$ and $h_\infty/x_0 = 0.0675$, both of which are $\ll1$ as required. The reduced Reynolds numbers are $(w/x_0)^2Re \approx 3 \times 10^{-3}$ and $(h_\infty/x_0)^2Re \approx 2.2 \times 10^{-2}$, both of which are $\ll1$ as required. {The corresponding Bond numbers are $Bo = \Delta\rho\,g w^2/\gamma \approx 3\times10^5$ and $(x_0/w)^2Bo = \Delta\rho\,g x_0^2/\gamma \approx 5\times10^8$, both of which are $\gg 1$ as required.}

Figure~\ref{fig:exptraw} shows a series of snapshots of the gravity current profile. The height of the current $h(x,t)$ was sampled intermittently at fixed instants of time and at discrete spatial locations $\{x_i\}_{i=1,2,\hdots}$. Then, this discretized shape was used to obtain the profile $H(x,\tau)$ via the transformation introduced in Eq.~\eqref{eq:h_H_transform}. Based on self-similarity analysis, it is expected that the rescaled experimental shape $H$, plotted against $\xi/\xi_f$, should agree with the second-kind self-similar profile computed from the ODE~\eqref{eq:HU_system3}. This agreement is, of course, contingent upon the gravity current having entered this intermediate asymptotic state. As discussed in the literature \cite{Diez1992}, the memory of the initial condition plays a significant role in the pre-closure regime. We now examine this issue in Section~\ref{sec:SSnum} below using numerical simulations of the governing lubrication PDE.

\subsection{Numerical study}
\label{sec:SSnum}

\subsubsection{Pre-closure self-similarity}
\label{sec:preclosure}
From the discussion in Section~\ref{sec:model} (see also \cite{Barenblatt1952,Gratton1990,zcs14}), it is expected that in second-kind self-similarity, $x_f(t)/x_f(0) \propto (\tau/t_c)^\delta$. From this expectation it follows that (recalling that $x_f(0)=x_0$):
\begin{equation}
    \frac{x_f(t)}{x_0} = \beta \left(\frac{\tau}{t_c}\right)^\delta \qquad\Longrightarrow\qquad \frac{x_f(t)}{\tau^\delta} = \frac{\beta x_0}{t_c^\delta}.
\label{eq:bintro}
\end{equation}
Here, $\beta$ is a `pre-factor' (proportionality constant) that must be obtained from numerical simulations and/or experiments. Importantly, the simulations and/or experiments must be accurate enough to determine whether the current has entered in the self-similar regime. Since $\xi_f = x_f/\tau^\delta$, by definition, it follows that 
\begin{equation}
    \xi_f = \frac{\beta x_0}{t_c^\delta},
    \label{eq:xif}
\end{equation}
where all terms on the right hand side are constant. Equation~\eqref{eq:xif} simply restates the assumption made in the self-similarity analysis that the similarity variable $\xi$ maintains the constant value $\xi_f$ at the current's front, $x=x_f$. It then follows that, during the initial adjustment from the initial condition, Eq.~\eqref{eq:xif} would not hold true. However, by $t=t_\mathrm{sim}$ (to be determined numerically), the adjustment would be complete, allowing for the pre-factor $\beta$ to be determined as the slope of curve generated by plotting $x_f(t)/x_0$ versus $({\tau}/{t_c})^\delta$.

\begin{figure}
\subfloat[][]{\includegraphics[width=0.49\textwidth]{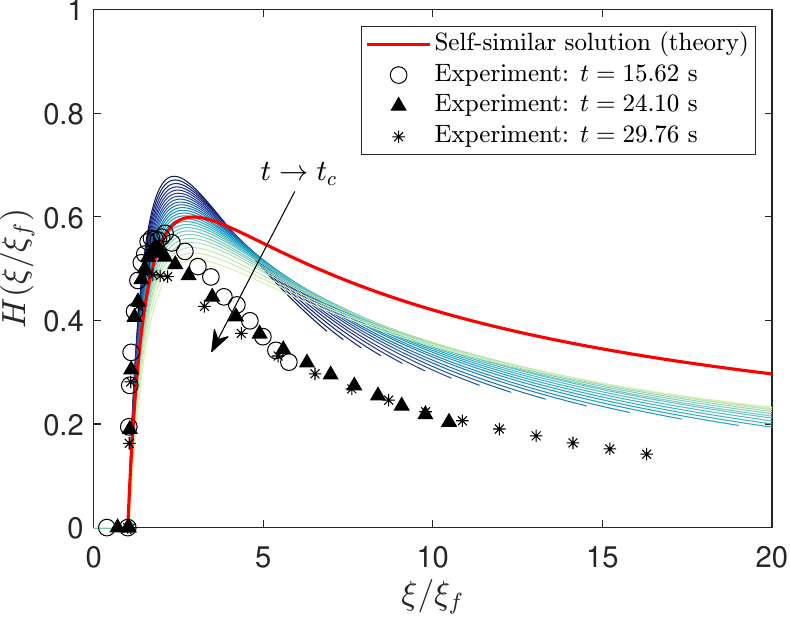}\label{fig:pc2}}
\hfill
\subfloat[][]{\includegraphics[width=0.485\textwidth]{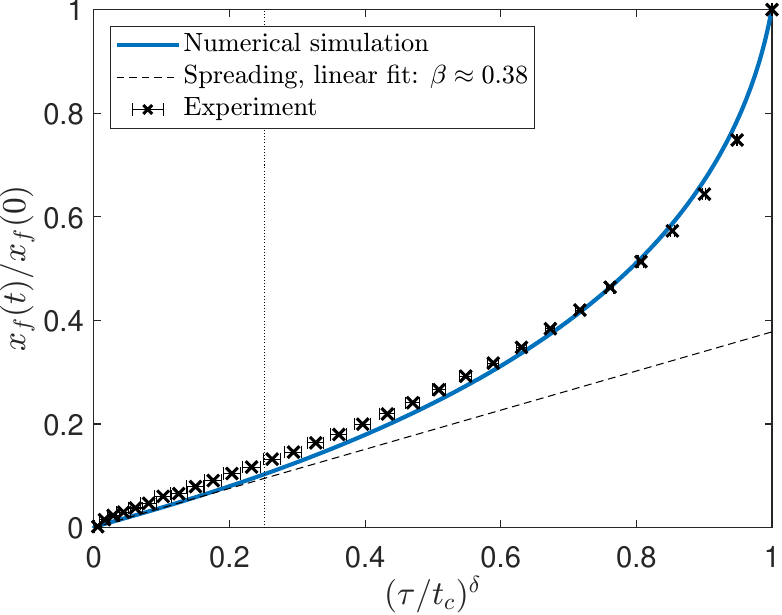}\label{fig:pc1}}
\caption{Study of the pre-closure (spreading) self-similar regime. The similarity exponent is $\delta \approx 1.5423$. (a) Comparison of the self-similar current shape profiles $H(\xi/\xi_f)$ between the predictions of self-similar theory, numerical simulation and experiment. (b) Determination of the pre-factor $\beta$, which sets $\xi_f$ via Eq.~\eqref{eq:xif}, and the time $t_\mathrm{sim} \approx 27.99$ s (denoted by vertical dotted line) after which the linear relationship holds. In (a), thin curves correspond to the rescaled profiles from numerical simulations of the governing lubrication PDE, color-coded by $t\in[t_\mathrm{sim},t_c]$ (increasing in the direction of the arrow) from dark to light (early times to late times). Error bars in (b) correspond to $t_c\pm 1$ s, based on the experimental technique employed.}
\label{fig:pre_closure}
\end{figure}

A numerical simulation matched to the experimental conditions described in Section~\ref{sec:expts} was performed and is shown in Fig.~\ref{fig:pre_closure}. The governing lubrication PDE for $h(x,t)$, which results from eliminating $u$ between Eqs.~\eqref{eq:udhdx} and \eqref{eq:conserv}, was solved numerically on the domain $x \in [0,L]$. {We employed a fully-implicit, second-order accurate (in space and time) finite-difference scheme specifically constructed for these types of variable-coefficient degenerate diffusion equations \cite{zcs14}. This finite-difference scheme uses a flux-conservative central discretization of the $x$-derivatives, and Crank--Nicolson-type time advancement, which circumvents the time-step stability restriction of explicit schemes. Fixed point (also known as Picard) iteration is used to resolve the nonlinear algebraic system at each time step. Importantly, this scheme captures the moving front (a weak discontinuity) without the requirement of unphysical `pre-wetting films' and so on, which were used in previous literature. Extensive benchmarking of the numerical method was reported in \cite{gc19}. 

The domain was discretized into $4501$ grid points. The simulation was run from $t_0 = 0$ s up to $t_f = 50$ s, over the course of $4501$ time steps. Grid and time step independence was verified. The initial condition $h(x,t=0)$ should ideally correspond to the shape of the profile just after the lock gate is opened. However, exact knowledge of this initial shape is difficult to obtain as the fluid shape adjusts quickly in the experiment (see also the discussion in \cite{Diez1992}). Instead, as a reasonable approximation, we selected a polynomial (satisfying the no-flux condition at $x=L$, see Eq.~\eqref{eq:scheme_noflux} below) as the initial condition for the simulations:
\begin{equation}
	h(x,0) = \begin{cases} a[(L-x_0)^c - (L-x)^c], &\quad x_0 \le x \le L,\\ 0, &\quad\text{otherwise}. \end{cases}
	\label{eq:scheme_ic}
\end{equation}
The dimensional constant $a$ is chosen to satisfy the volume constraint $\int_{x_0}^L h(x,0)b(x)\,\mathrm{d}x=\mathcal{V}$. For $c\ge 3$, the effect of $c$ on the propagation results was found to be negligible, hence we employ $c=3$, which yields a `boxy' shape for the initial gravity current. The reader is referred to Refs.~\cite{gc19,AAG19} for further discussion and validation of this choice of initial condition. To enforce the volume constraint, no-flux BCs are applied at the ends of the computational domain:
\begin{equation}
	\left.\left[ b(x) h u\right]\right|_{x\to0} = \left.\left[ b(x) h u\right]\right|_{x=L} = 0
	\quad \Rightarrow\quad \left.\left( x^{3n} h \frac{\partial h}{\partial x}\right)\right|_{x\to0} = \left.\left( x^{3n} h \frac{\partial h}{\partial x}\right)\right|_{x=L} = 0.
	\label{eq:scheme_noflux}
\end{equation}
See Refs.~\cite{zcs14,gc19} for discussion of the numerical implementation of these BCs.
} 

The simulations yielded a closure time of $t_c = 49.38$ s, showing reasonable agreement with the experimental measurement, and justifying the choice of initial condition for the simulations. Moreover, although the choice of initial condition is also expected to have an effect on the value of $t_\mathrm{sim}$, which roughly represents the time required for the solution to enter the intermediate asymptotic self-similar regime, our numerical experiments suggest only a weak dependence. As discussed in Appendix~\ref{app:B}, the BCs (no-flux versus influx) have much a stronger effect.

Figure~\ref{fig:pre_closure}\subref{fig:pc2} compares the self-similar gravity current shape $H(\xi/\xi_f)$ as obtained from theory (i.e., the solution of Eq.~\eqref{eq:ppode} described in Section~\ref{sec:ODE_solve}) to numerical simulation of the governing lubrication PDE and experimental measurements. The numerical profiles from the PDE were rescaled using the transformation in Eq.~\eqref{eq:h_H_transform} in the time period $t \in [t_\mathrm{sim},t_c]$. The theoretical, rescaled numerical, and rescaled experimental profiles show collapse near the front, $\xi/\xi_f=1$. {Note that, since the self-similarity here is a local concept, good agreement appears in a region near the front ($\xi/\xi_f=1$), and before the $H(\xi/\xi_f)$ profile shape reaches its maximum. Disagreement starts to appear beyond the maximum point of the rescaled profile.}

As shown in Fig.~\ref{fig:pre_closure}\subref{fig:pc1}, the linear proportionality between $x_f(t)/x_0$ and $(\tau/t_c)^\delta$ holds well for $t > t_\mathrm{sim} \approx 27.99$ s. This observation is made by first performing a linear fit on $x_f(t)/x_0$ values from the last 600 time steps of the simulation before closure to yield the pre-factor value of $\beta \approx 0.38$. Then, we work `backwards,' comparing the local value of $x_f(t)/x_0$ at each scaled time $(\tau/t_c)^\delta$ to the linear fit $\beta (\tau/t_c)^\delta$ up to that instant of time. Finally, $t_\mathrm{sim}$ is taken to be the threshold at which the local value of the curve disagrees with the linear fit by more than $\approx 1\%$. This analysis suggest that the time interval over which the intermediate self-similar asymptotics hold should be approximately $[t_\mathrm{sim},t_c]$. 
Appendix~\ref{app:B} provides further discussion on the possible reasons for any disagreement observed in the rescaled profiles shown in Fig.~\ref{fig:pre_closure}\subref{fig:pc2}.

\subsubsection{Post-closure self-similarity}
\label{sec:post_closure}

The theory of the post-closure (or, leveling) self-similar regime was established in Section~\ref{sec:model}. However, unlike during pre-closure, the position of the current's front $x_f(t)$ during post-closure is fixed; specifically, it remains at the origin, i.e., $x_f(t>t_c) = 0$. This fact necessitates the replacement of $x_f(t)$ as a dynamic length scale. To this end, the height of the current at the origin of the channel, $h(0,t) >0$, is now justified as the dynamic length scale.

The post-closure self-similar solution corresponds to the integral curve connecting Point D to Point O in the phase plane (recall Fig.~\ref{fig:pplane}), and this curve's asymptotics near Point D were calculated in Section~\ref{sec:pointD}. To obtain an expression for $h(0,t)$, we observe that Eq.~\eqref{eq:hxt_tip_post} has a well defined limit as $x\to0^+$, yielding:
\begin{equation}
h(0,t) \sim \left(\frac{12\mu}{\Delta\rho g b_1^2}\right) \frac{\zeta_f^{2(1-n)}}{(-\tau)^{1-2\delta(1-n)}} \qquad\text{for}\qquad h\to0,
\label{eq:h0t}
\end{equation}
keeping in mind that $\tau<0$ ($t > t_c$) in the post-closure regime. Introducing $h_\infty = \lim_{t\to\infty}h(0,t)$, from Eq.~\eqref{eq:h0t}, we deduce that self-similarity requires
\begin{equation}
	\frac{h(0,t)}{h_\infty} \sim  \underbrace{\left(\frac{12\mu}{\Delta\rho g b_1^2}\right) \frac{\zeta_f^{2(1-n)}t_c^{2\delta(1-n)-1}}{h_\infty}}_{\beta} (-\tau/t_c)^{2\delta(1-n)-1}.
	\label{eq:h0scaling}
\end{equation}
The pre-factor $\beta$ in Eq.~\eqref{eq:h0scaling} can thus be used to determine $\zeta_f$ in post-closure from a linear fit of $h(0,t)/h_\infty$ vs.\ $(-\tau/t_c)^{2\delta(1-n)-1}$. Observe that $2\delta(1-n)-1 \equiv (n+1)U_\mathrm{D}$.

In Eq.~\eqref{eq:h0scaling}, $h_\infty$ represented the steady-state height of the current. Based on conservation of mass, it is easy to show that 
\begin{equation}
h_\infty = \frac{\int_0^L b_1 x^n h(x,0^+) \,\mathrm{d} x}{\int_0^L b_1 x^n\, \mathrm{d}x}.
\label{eq:hinf}
\end{equation}
In Eq.~\eqref{eq:hinf}, the numerator is equal to the total volume $\mathcal{V}$ of fluid released, and the denominator represents the horizontal cross-sectional area of the variable-width HS cell. From the parameters in Table~\ref{tb:expt}, we obtain $h_\infty \approx 0.0333$ m.

\begin{figure}[t]
\subfloat[][]{\includegraphics[height=0.36\textwidth]{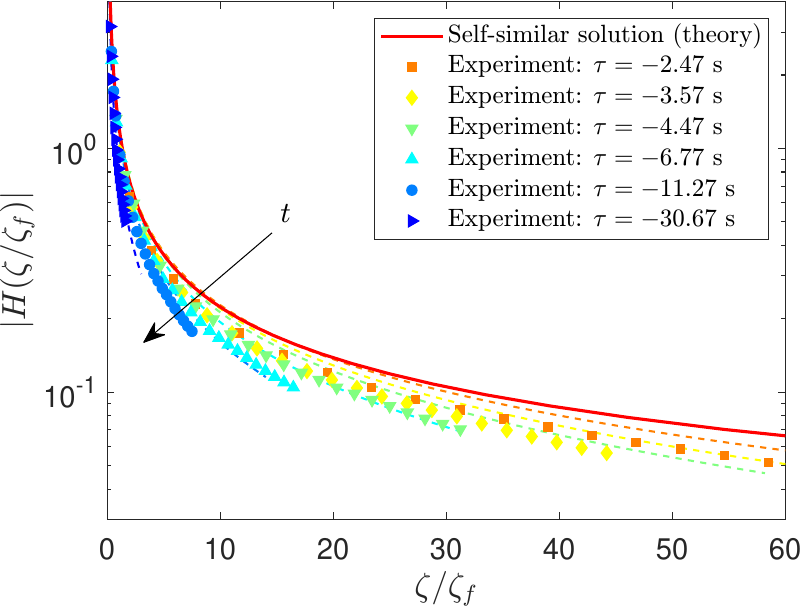}\label{fig:post2}}
\hfill
\subfloat[][]{\includegraphics[height=0.365\textwidth]{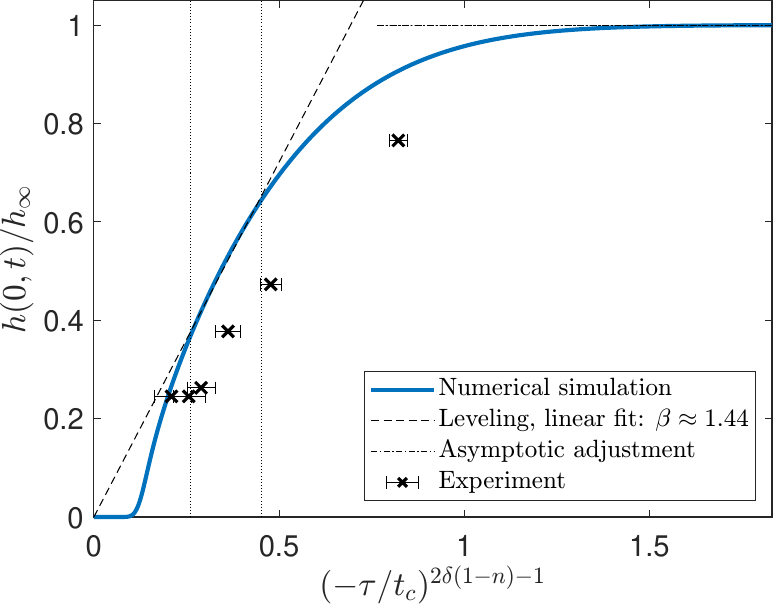}\label{fig:postc1}}
\caption{Study of the post-closure (leveling) self-similar regime. The similarity exponent is now $2\delta(1-n)-1 \approx 0.5423$. (a) Comparison of the self-similar current shape profiles $H(\zeta/\zeta_f)$ between the predictions of self-similar theory, numerical simulation and experiment. (b) Determination of the pre-factor $\beta$, which sets $\zeta_f$ via Eq.~\eqref{eq:h0scaling} (needed to rescale the simulation data in (a)), and the times $t_\mathrm{sim}^{(1)}\approx 53.5$ s and $t_\mathrm{sim}^{(2)} \approx 61$ s (denoted by vertical dotted lines) between which the linear relationship holds. The arrow in (a) indicates the direction of increasing $t$; note that the vertical scale is logarithmic to better highlight the shapes. Dashed curves in (a) are from numerical simulations of the governing lubrication PDE, matched to the $\tau$ values of the experiments (symbols). Error bars in (b) correspond to $t_c\pm 1$ s, based on the experimental technique employed.}
\label{fig:post_closure}
\end{figure}

As can be seen in Fig.~\ref{fig:post_closure}\subref{fig:post2}, during the leveling period, excellent agreement is observed between the self-similar gravity current profiles, $H(\zeta/\zeta_f)$, obtained from theory, numerical simulation and the experiment. The universal profile from the theory established in Section~\ref{sec:model} was obtained, once again, by solving the ODE system~\eqref{eq:HU_system3}. The numerical profiles computed for $h(x,t)$ from the simulation of the governing lubrication PDE were scaled via the self-similarity transformation in Eq.~\eqref{eq:h_H_transform} and shown for the same values of $\tau$ (which is independent of the $5.36$~s difference between simulated and measured $t_c$ values). The experimental data comes from digitally sampling images (Fig.~\ref{fig:exptraw}) of the profile $h(x,t)$ at six distinct times in the post-closure (i.e., $t> t_c$ and $\tau <0$) regime. It should be noted that, unlike the axisymmetric converging current \cite{Diez1992}, the converging current in the variable-width HS cell is not difficult to image post-closure, and the overall profile shapes are captured well, as seen in Fig.~\ref{fig:post_closure}\subref{fig:post2}. The good agreement confirms that second-kind self-similarity exists in the post-closure regime. 

In post-closure, the memory of the current's initial (or boundary) condition used in the simulation no longer have an effect, as the self-similar process `resets' after the current reaches the origin. However, the governing lubrication PDE, having neglected capillary effects, which are important near the origin, undergoes another adjustment period (reminiscent of the so-called `waiting-time' solution of nonlinear parabolic PDE \cite{Lacey1982}, which is a type of phase-plane integral curve not discussed here) before entering the post-closure self-similar regime. The trend of $h(0,t)/h_\infty$ versus $(-\tau/t_c)^{2\delta(1-n)-1}$ in Fig.~\ref{fig:post_closure}\subref{fig:postc1} indicates self-similarity as the current levels during post-closure. However, the stretched horizontal (time) coordinate overemphasizes the earliest $\tau$ at which the values of $h(0,t)$ are barely distinguishable from zero. Therefore, it is most logical to interpret the `middle' portion of the simulation curve in Fig.~\ref{fig:post_closure}\subref{fig:postc1} as representing the linear relationship predicted by self-similarity. Indeed, a good fit is obtained by fitting the `middle third' range of values such that $1/3\le h(0,t)/h_\infty\le 2/3$, yielding the pre-factor $\beta \approx 1.44$. Testing a number of possible ranges for the fit revealed that the value of $\beta$ is not particularly sensitive to the chosen bounds of $1/3$ and $2/3$. Indeed, it is expected that the post-closure second-kind self-similarity takes a finite amount of time to establish itself \cite{Diez1992}, providing further support for our intermediate fit. Finally, we observe that the experimental measurements of $h(0,t)/h_\infty$ shown in Fig.~\ref{fig:post_closure}\subref{fig:postc1} are not as accurate as the experimental measurements of $x_f(t)/x_0$ in Fig.~\ref{fig:pre_closure}\subref{fig:pc1}. 

Thus, we have motivated that the predicted post-closure self-similarity is indeed an intermediate asymptotic state that is self-consistently manifested in this system. However, this state may not be the only self-similar post-closure regime. In addition to the apparent `waiting-time' behavior near $\tau=0$, after some time ($> 0.5$ dimensionless units in Fig.~\ref{fig:post_closure}\subref{fig:postc1}), a transition begins from leveling towards an asymptotic adjustment as $h(0,t) \to h_\infty$. This latter regime is, however, a relatively straightforward observation, and it is not of further interest here.


\section{Conclusion}

Self-similar transformations have been shown to be a powerful tool for analyzing fluid mechanical phenomena \cite{Sedov1993,Barenblatt1979,Eggers2015}. Although not all viscous gravity currents are self-similar \cite{Sutherland2018}, first-kind self-similar solutions can be obtained through a scaling analysis of a suitable mathematical model. By seeking a first-kind self-similar solution, it is possible to reduce the current's governing PDE to an (often) exactly solvable ODE (in closed form, again see e.g., Refs.~\cite{Barenblatt1952,Huppert1995,Anderson2003,Lyle2005,Vella2006,Hesse2007,Anderson2010,DeLoubens2011,Ciriello2013,Zheng2013,zcs14,zfccs15,Furtak-Cole2018,McCue2019,Sahu2020} as well the review-style discussions \cite{Longo2015,Ciriello2016}). Meanwhile, in flow regimes involving additional spatial (or temporal) scales, a scaling analysis is insufficient to reduce the governing PDE to a closed-form self-similar solution. Instead the problem requires, for example, using phase-plane analysis and solving a nonlinear eigenvalue problem to determine the so-called `anomalous exponents' \cite{Aronson1994}, which allow second-kind self-similar transformations to be specified. However, conducting additional numerical simulations (and/or experiments) is also necessary to determine certain numerical constants (pre-factors) that are arbitrary in the second-kind self-similar transformation. 

In this study, second-kind self-similarity was explored in the context of the release of a fixed mass of Newtonian fluid spreading towards the origin of a horizontal, shaped Hele-Shaw cell of variable width. A self-similar transformation was introduced, depending upon an extra time scale (arising from the time it takes the current to reach the closed end of the cell) and also upon an anomalous exponent $\delta$ (found numerically as an eigenvalue). Importantly, a detailed phase-plane analysis was conducted to extend the results from \cite{zcs14} and to provide novel asymptotics for the integral curves (and, consequently, closed-form approximations for the gravity current's shape) for both the pre- and post-closure self-similar regimes. In parallel, numerical simulations of the governing lubrication theory PDE allowed us to determine whether (and how) such flow regimes, predicted by mathematical theory, are actually manifested during the spreading and leveling of gravity currents in variable-width channels.

However, in determining whether a given gravity current profile $h(x,t)$, found from either simulation or experiment, will collapse onto the predicted universal profile $H$, obtained by second-kind self-similarity analysis, knowledge of the precise time period during which the current is expected to be in this self-similar regime was needed. Previous work had not addressed the question of when self-similarity `begins,' and it is indeed a difficult mathematical question in general. We proposed an approximation to this time, $t_\mathrm{sim}$, from numerical simulation of the governing lubrication PDE by tracking a suitable dimensionless spatial scale against a predicted power of the dimensionless time-to-closure.

In summary, this combined theoretical--numerical--experimental study demonstrates the presence of two distinct self-similar regimes of the second-kind occurring, respectively, during (i) spreading (pre-closure) and then (ii) leveling (post-closure) of a gravity current in a variable-width channel, which is a canonical flow configuration. The second-kind self-similar asymptotics derived generally hold only near the current's moving front (`nose'). Consequently, we only observed qualitative agreement between the experimental gravity current shape profiles away from the front, during pre-closure, and the theoretical and numerical simulations for this case of constant volume. During post-closure, however, the experimental gravity current profiles, scaled via the second-kind self-similar transformation, showed good agreement with the theoretical predictions and numerical simulations.
{
Further, in comparison to the contemporaneous study by \citet{Longo2021}, the following differences should be noted.
\begin{enumerate}
	\item We provided numerical solutions of the governing lubrication theory PDE, unlike Ref.~\cite{Longo2021}. The simulations turn out to be critical in bridging the experimental measurements with the mathematical theory of the second-kind self-similarity.
	\item We demonstrated that the self-similarity of the current profile is only manifested during an \emph{intermediate} time period. That is, self-similarity takes a finite amount of time to establish itself. This issue, although mentioned very briefly on page 23 in Ref.~\cite{Longo2021} was not addressed, perhaps due to limitations of the experiments and/or lack of requisite numerical solutions of the governing lubrication theory PDE. This issue is particularly important in the post-closure regime (recall Fig.~\ref{fig:pre_closure}(b) above and its discussion).
	\item We related the undetermined scaling pre-factors (pre- and post-closure) to physical quantities in Eqs.~\eqref{eq:bintro} and \eqref{eq:h0scaling}, rather than taking them as fitting constants (as in Ref.~\cite{Longo2021}). We showed that the pre-factor $\beta$ depends on geometric and physical quantities, and that it can be inferred for a given initial condition from numerical solutions of the governing lubrication theory PDE. The key point is that there remains a dimensional constant $\zeta_f$, which depends on the initial (and boundary) conditions, that must be determined from simulations. Under second-kind self-similarity, despite there being \emph{universal} behaviors, the current is not truly `memoryless,' contrary to the statements in Ref.~\cite{Longo2021}, as the self-similarity is `incomplete' \cite{Barenblatt1972,Barenblatt1979}.
\end{enumerate}
}
	
In future work, it would be of interest to address, via new experiments and detailed simulations, the collapse of gravity current profiles in the pre-closure regime. One possibility is to consider a time-dependent injection rate (see the discussion in Appendix~\ref{app:B}). The reasoning is that, the self-similar theory presented is local, being valid when the front is close to the dead-end, which means that the global volume conservation (upon transforming $H(\xi/\xi_f)$ back to $h(x,t)$) includes a time dependence. Furthermore, a time-dependent injection rate is likely to better satisfy the quasi-steady-state assumption on the self-similar solution's far-field behavior, which is implicit in the self-similar transformation \cite{Wang1999,ZhengHealing}. Therefore, in future work, it would be appropriate to compare the theory to an experiment with a suitable influx at $x=L$, instead of the condition of constant volume used in our current experiments.


\section*{Acknowledgments}
We would like to thank Howard A.\ Stone for fruitful discussions on self-similarity and his input. A preliminary version of some portions of this work appeared in A.A.G.'s Masters thesis \cite{AAG19}.


\bibliography{2nd_kind_references}


\clearpage
\appendix

\section{Center manifold reduction near Point O}
\label{app:pointO}

Finding the asymptotic behavior for integral curves near Point O requires some care due to the degeneracy (zero eigenvalue) \footnote{In dynamical systems, a fixed or singular point with one zero and one negative eigenvalue is considered a ``marginal case,'' meaning the linearization about about this point may or may not provide its \emph{stability}. But stability is not our concern here as we easily infer the dynamics from the numerical phase plane plot.}. Observe from Fig.~\ref{fig:pplane} that the slope of integral curves near Point O is the same, whether arriving along the heteroclinic from Point D \emph{or} leaving on the heteroclinic to Point A. We know that the unstable manifold at Point O has slope \cite[Eq.~(2.16{\it b})]{zcs14}:
\begin{equation}
\frac{\mathrm{d}U}{\mathrm{d}H} \sim -\frac{2(1-n)\delta-1}{\delta} \quad\Rightarrow\quad U \sim -\left[\frac{2(1-n)\delta-1}{\delta}\right] H,
\label{eq:slope_OD}
\end{equation}
where the constant of integration is set by requiring that the line goes through Point O. 

Then, we decompose the autonomous system~\eqref{eq:HU_system} into linear and nonlinear parts:
\begin{equation}
\frac{\mathrm{d}}{\mathrm{d}s}\begin{pmatrix} H\\ U \end{pmatrix} 
= \underbrace{\begin{pmatrix} 0\\ [1-2(1-n)\delta]H - U\delta \end{pmatrix}}_\text{linear} + \underbrace{\begin{pmatrix} H[2(1-n)H + U]\\ (n+1)HU - U^2 \end{pmatrix}}_\text{nonlinear}.
\label{eq:HU_system2}
\end{equation}
Clearly, along the line defined in Eq.~\eqref{eq:slope_OD}, the dynamical system in Eq.~\eqref{eq:HU_system2} is \emph{genuinely} nonlinear, making it a candidate for a center manifold reduction \cite[\S3.2]{Guckenheimer1983}. To find the center manifold, suppose $U$ can be expressed as a polynomial $P$ in $H$, i.e.,
\begin{equation}
U = P(H) = a_1 H + a_2 H^2 + a_3 H^3 + \cdots, \qquad\text{near}\;\; \mathrm{O},
\label{eq:cm_exp}
\end{equation}
where $a_1$ is known from Eq.~\eqref{eq:slope_OD} above. Then,
\begin{subequations}\begin{align}
\frac{\mathrm{d}U}{\mathrm{d}s} &= H[(n+1)P - 2(1-n)\delta + 1] - P(P+\delta),
\label{eq:dUds_center}\\
\frac{\mathrm{d}P}{\mathrm{d}s} &= \frac{\mathrm{d}P}{\mathrm{d}H}\frac{\mathrm{d}H}{\mathrm{d}s} = \frac{\mathrm{d}P}{\mathrm{d}H} H[2(1-n)H + P].
\label{eq:dUds_center2}
\end{align}\end{subequations}
Eqs.~\eqref{eq:dUds_center} and \eqref{eq:dUds_center2} together define a first-order ODE for $P(H)$:
\begin{equation}
\frac{\mathrm{d}P}{\mathrm{d}H} H[2(1-n)H + P] = H[(n+1)P - 2(1-n)\delta + 1] - P(P+\delta).
\label{eq:dPdH}
\end{equation}
Substituting the expansion from Eq.~\eqref{eq:cm_exp} into Eq.~\eqref{eq:dPdH} and keeping only terms up to $H^3$, we obtain
\begin{multline}
a_1[a_1 + 2(1-n)]H^2 + \left\{ 2a_2[a_1 + 2(1-n)] + a_1a_2 \right\} H^3 + \cdots\\
 = \underbrace{[1-2(1-n)\delta - \delta a_1]}_{=0\text{ by definition of }a_1} H + [(n+1)a_1-\delta a_2 - a_1^2] H^2 + [(n+1) a_2 - \delta a_3 - 2a_1a_2] H^3 + \cdots.
 \label{eq:on_CM}
\end{multline}
Equating the coefficients of $H^2$ and $H^3$ on each side of Eq.~\eqref{eq:on_CM}:
\begin{subequations}\begin{align}
a_2 &= -\frac{(1 + 2 a_1 - 3 n)a_1}{\delta} = \frac{[(n-3)\delta+2][2(1-n)\delta-1]}{\delta ^3},\\
a_3 &= -\frac{(5 a_1 - 5n + 3)a_2}{\delta} = -\frac{[(n-3)\delta+2][2(1-n)\delta-1][(5 n-7)\delta+5]}{\delta ^5} .
\end{align}\end{subequations}
In principle, using a computer algebra system, one can go to even higher orders, obtaining successive corrections. As we can see from the black curve near marked `center manifold' in Fig.~\ref{fig:pplane}, the polynomial center manifold reduction improves the accuracy of the linear approximation, i.e., Eq.~\eqref{eq:slope_OD}.

Finally, following the earlier procedure (from Sections~\ref{sec:pointA} and \ref{sec:pointD}), the center manifold from Eq.~\eqref{eq:cm_exp} (keeping only two terms) can be rewritten as 
\begin{equation}
\frac{\partial h}{\partial x} \sim -a_1 \frac{1}{x}h -a_2 \left(\frac{\Delta\rho g b_1^2}{12\mu}\right)\frac{\tau^2}{x^{3-2n}} h^2 + \cdots,
\end{equation}
which integrates to
\begin{equation}
h(x,t) \sim \left\{K x^{a_1} - \left(\frac{\Delta\rho g b_1^2}{12\mu}\right) \left( \frac{a_2 \tau^2 x^{-2(1-n)}}{2a_1-2n} \right) \right\}^{-1}.
\end{equation}
The constant of integration $K$ cannot be determined from this analysis, and a full numerical simulation (or experiment) must be performed to determine $K$ for some chosen initial conditions.


\clearpage
\section{Discussion of disagreement observed in rescaled pre-closure profiles}
\label{app:B}

Our phase-plane analysis from Sections~\ref{sec:model} and \ref{sec:shape_calc} focused on the local dynamics of the pre-closure self-similarity, based on connecting Point A (moving front) with Point O (origin/closed end of the HS cell). The so-called `far-field behavior' (as $\xi\to\infty$) \cite{Wang1999,ZhengHealing} of the second-kind self-similar solution was not analyzed. Indeed, it is not hard to check numerically that, when mapped from $H(\xi/\xi_f)$ back to $h(x,t)$, the self-similar solution (found by numerically integrating the phase plane ODE~\eqref{eq:HU_system3}) does not maintain a constant volume on $x\in[0,L]$. However, our experimental study was performed under the condition of constant volume release. To highlight the potential effect of this discrepancy, we performed numerical simulations {replacing the no-flux BC at $x=L$ in Eq.~\eqref{eq:scheme_noflux} with a BC that corresponds to an injection rate:
\begin{equation}
\left.[b(x)hu]\right|_{x\to0} = 0,\qquad \left.[b(x)hu]\right|_{x=L} = \dot{\mathcal{V}}_{\text{in}},
\end{equation}
where $\dot{\mathcal{V}}_{\text{in}}$ is the volumetric inflow \emph{rate} at $x=L$ such that $\dot{\mathcal{V}}_{\text{in}}=\mathrm{d}\mathcal{V}/\mathrm{d}t$.
For these simulations, the initial condition was taken to be the self-similar solution found by numerically integrating the phase plane ODE~\eqref{eq:HU_system3}, and mapped back to $t=0$ with $t_c =44.03$~s (per the experiments) and $\beta = 0.38$ (per Fig.~\ref{fig:pre_closure}\subref{fig:pc1}), instead of Eq.~\eqref{eq:scheme_ic}.} Note that this initial condition sets a $\mathcal{V}_0$ different from the experimental value based on Table~\ref{tb:expt}. Then, we tuned a constant injection rate $Q_\mathrm{in}$, such that $\mathcal{V}=\mathcal{V}(t) = \mathcal{V}_0 + Q_{\text{in}}t$, until $t_c$ from the simulation matched the experimental value $44.03$~s used to set the initial condition. However, $Q_\mathrm{in}$ is not the instantaneous volume change of the second-kind self-similar solution, which although slowly varying, is not constant.

\begin{figure}
\subfloat[][]{\includegraphics[width=0.49\textwidth]{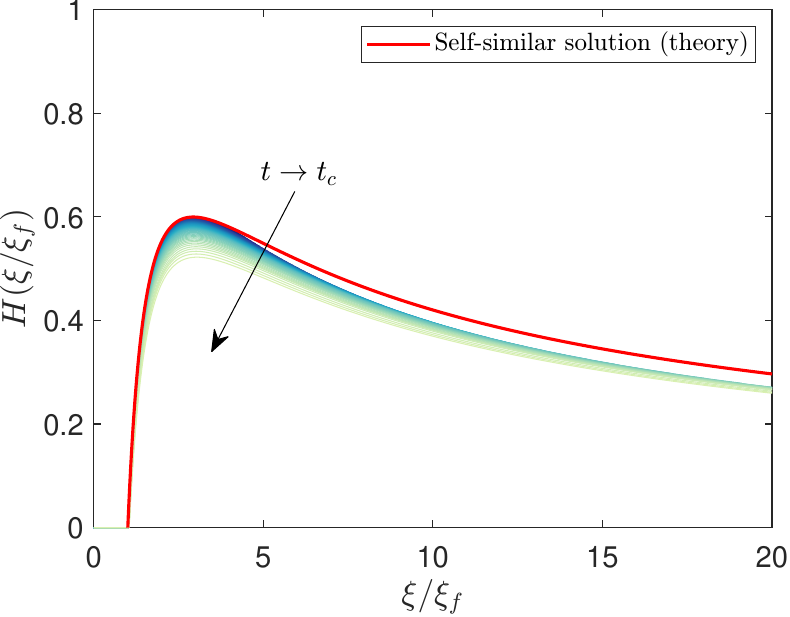}\label{fig:pc2i}}
\hfill
\subfloat[][]{\includegraphics[width=0.485\textwidth]{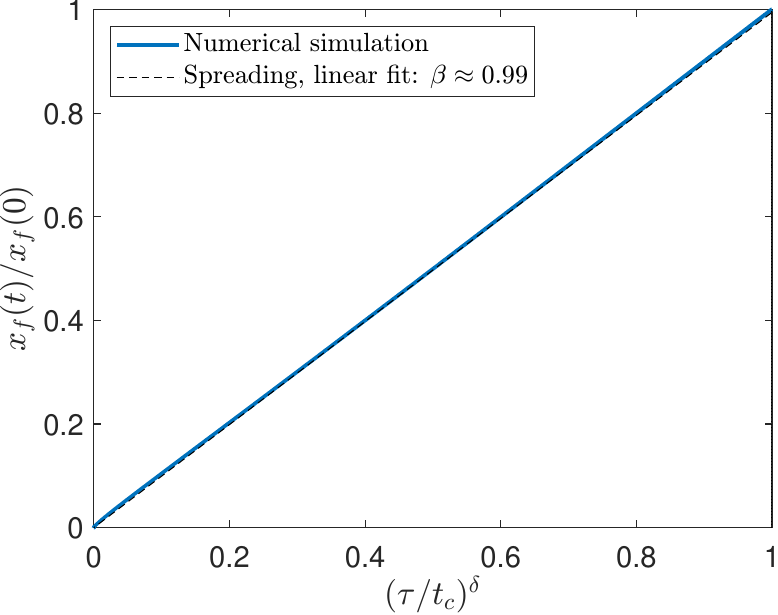}\label{fig:pc1i}}
\caption{Study of the pre-closure (spreading) self-similar regime with inflow at $x=L$ to maintain $\mathcal{V}(t) = \mathcal{V}_0 + Q_{\text{in}}t$ with $Q_\mathrm{in} \approx 5.025 \times 10^{-6}$ m\textsuperscript{3}/s. (a) Comparison of the self-similar current shape profiles $H(\xi/\xi_f)$ between the predictions of self-similar theory and numerical simulation with influx at $x=L$. (b) 
Having used the self-similar solution as an initial condition, and imposed a suitable influx at $x=L$, the pre-factor $\beta\approx 1$ and $t_\mathrm{sim} \approx 0$ s. In (a), thin curves correspond to the rescaled profiles from numerical simulations of the governing lubrication PDE, color-coded by $t\in[t_\mathrm{sim},t_c]$ (increasing in the direction of the arrow) from dark to light (early times to late times).}
\label{fig:pre_influx}
\end{figure}

The results are shown in Fig.~\ref{fig:pre_influx} for the `tuned' value of $Q_{\text{in}} \approx 5.025 \times 10^{-6}$ m\textsuperscript{3}/s. Clearly, the rescaled $h(x,t)$ profiles from numerical simulation in Fig.~\ref{fig:pre_influx}\subref{fig:pc2i} collapse onto the universal theoretical shape $H(\xi/\xi_f)$ far better than in Fig.~\ref{fig:pre_closure}\subref{fig:pc2}, suggesting that the `far-field' condition required by second-kind self-similarity has nontrivial consequences. Unsurprisingly, the fitted value of $\beta\approx 1$ in Fig.~\ref{fig:pre_influx}\subref{fig:pc1i} is different from Fig.~\ref{fig:pre_closure}\subref{fig:pc1} because this pre-factor depends, as discussed above, on the initial condition, which in this simulation was taken to be the self-similar solution itself. A consequence of the chosen initial condition, together with the influx BC used for these simulations, is that the interval $[t_\mathrm{sim},t_c]$, over which self-similarity is expected to hold, increases to the full time interval of the simulation $[0,t_c]$. Since the influx condition here is only approximate, the late-time (light-color) profiles in Fig.~\ref{fig:pre_influx}\subref{fig:pc2i} eventually `drift' away from the theory curve. 

The key conclusion from this numerical exploration is that the second-kind self-similar pre-closure solution implies an influx condition, which is different from the particular experiments discussed in the main text above. {It is also worth noting, consistent with the discussion in this appendix, that a recent experimental study concluded that ``achievement of the self-similar condition is faster for constant inflow rate than for lock release'' (constant volume) \cite[p.~25]{Longo2021}.}

\end{document}